\documentstyle[psfig,times]{mn}
\title[Gravitational radiation from highly magnetized neutron stars]{Gravitational radiation from highly magnetized nascent neutron stars in supernova remnants}
\author[Shin Yoshida]{Shin'ichirou Yoshida \footnote{yoshida@sissa.it} \\
	SISSA / International School for Advanced Studies, Via Beirut 2-4, 34014 Trieste, Italy}
\date{}
\pagerange{\pageref{firstpage}--\pageref{lastpage}}
\pubyear{2002}

\begin{document}

\maketitle

\begin{abstract}
	We consider the spin evolution of highly magnetized neutron 
	stars in a hypercritical inflow just after their 
	birth in supernovae.
	Presence of a strong magnetic field could deform the star and
	if the symmetry axis of the field is misaligned with that of
	stellar rotation, the star will be an emitter of gravitational wave.
	Here we investigate the possibility of gravitational radiation
	from such a star when there is a hypercritical inflow onto it.
	For doing this we adopt a simplified model of the system in which 
	the star is approximated as a Newtonian spherical polytrope 
	with index $N=1$.
	The stellar configuration is slightly deformed away from spherical
	by the intense magnetic field; the rotational angular 
	frequency of the star is determined by the balance between
	the accretion torque and the magnetic dipole radiation.
	We take into account the 'propeller' process in which
	a rotating stellar magnetic field flings away in-falling matter;
	the inflow is assumed to be a self-similar advection
	dominated flow.
	An estimation of the characteristic amplitude of the gravitational
	radiation from such systems is given. The computation of the
	signal-to-noise ratio suggests that for the case of 
	an initially rapidly rotating
	and highly magnetized star (surface field $10^{15}$G) in
	the Virgo Cluster,
	its ellipticity would need to be larger than $10^{-5}$ in order for
	the gravitational waves to be observed.
\end{abstract}

\section{Introduction}

Recently Watts \& Andersson (2002) have considered the possibility
of detecting gravitational radiation from a newly born neutron star 
in a supernova
remnant. In their scenario, debris of a supernova explosion (typically 
$\sim 0.1M_\odot$) falls back hypercritically ($\dot{M}>10^{-4}M_\odot \mbox{yr}^{-1}$) 
onto a neutron star with a normal-strength magnetic field ($B<10^{13}$G)
and transfers angular momentum to the star. 
On the other hand, the r-mode instability\footnote{
See Andersson \& Kokkotas (2001) for a review of this instability and
its astrophysical implications.} which is driven by gravitational radiation,
can extract angular momentum from the system quite efficiently.
The accretion torque and the torque of gravitational radiation
balance to give the star a characteristic rotational period of a few 
milliseconds. As a result, this system can be an efficient emitter
of gravitational waves of nearly constant frequency for several years.
\footnote{Yoshida \& Eriguchi (2000) have considered the r-mode
instability of a hypercritically accreting neutron star in a common 
envelope with a normal star companion.}
Their work indicates that for stars with a normal-strength magnetic 
field ($B<10^{13}$G), the system can be a promising source for
gravitational wave.

On the other hand, for moderately magnetized stars 
(i.e., $B\sim 10^{10}-10^{12}$G) Rezzolla et al. 
have argued that the coupling of r-modes with the 
magnetic field can prevent the instability from developing.
Thus the results by Watts \& Andersson may not be relevant to the stars
with rather strong magnetic fields ($B>10^{13}$G)

As a complementary study to theirs, we study here a similar system 
but with the neutron star having a much larger magnetic field
($B\sim 10^{14}-10^{15}$G). 
Magnetic fields of this strength are associated with the so-called
{\it magnetars} (Duncan \& Thompson 1996). 

When the magnetic field reaches such large intensities,
the equilibrium configuration would be distorted by the magnetic tension
considerably.
In the case of an axisymmetric magnetic field, the star will be deformed
in an axisymmetric way aligned with the axis of the magnetic field 
(Bonazzola and 
Gourgoulhon 1996).
If this symmetry axis is not aligned with the rotational one,
this will produce a time varying quadrupole moment and
lead to the emission of gravitational waves.

In this respect, our system is quite different from the one considered by
Watts \& Andersson (2002), since here gravitational radiation is
not produced by the r-mode instability, and has only
a passive role in the evolution of the system. The evolution of
the system, i.e., of the rotational angular frequency and 
of the mass, is regulated by the evolution of the accretion torque.

In particular, the so-called 'propeller' mechanism (Pringle \& Rees 1972,
Illarionov \& Sunyaev 1975) becomes active rather early.
(This mechanism consists of the in-falling matter being flung away 
from the stellar magnetosphere when the rotational frequency is larger 
than the local orbital frequency, because the centrifugal force on the 
matter exceeds the gravitational force.) 
Also the torque from magnetic dipole radiation 
may be important in the later phases for highly magnetized stars.

The plan of the paper is the following. In section 2, the formulation of 
our model is outlined. In the following section, the results,
(i.e., the evolution of the stellar rotational frequency and the mass, 
the characteristic amplitudes of gravitational waves from the system,
and the signal-to-noise ratio of gravitational wave for 
laser-interferometric detectors), are presented for several parameter sets. 
The final section contains the summary and some comments about our model.

\section{A Simplified model}
Our model of the system is rather simplified
and follows closely the one presented by Watts \& Andersson (2002).
Our description of the physical processes is rather
idealized but, nevertheless, 
it allows the basic features of the gravitational wave emission
from these systems to be appreciated.

The star is assumed to be a Newtonian polytrope with polytropic
index $N=1$, having radius $R$ and the mass $M$. 
Debris of the supernova falls back onto the star according to a
simple power law of time, $\dot{M}_{_{fall}}\sim t^{-3/2}$
(Mineshige et al. 1997; Menou et al. 1999). 
Note that this in general should be distinguished from the
mass accretion rate onto the star $\dot{M}$, especially 
when the propeller mechanism is active. We assume that,

\begin{eqnarray}
	\dot{M} &=& \dot{M}_{_{fall}}	\qquad \mbox{(propeller is inactive)}\nonumber\\
		&=& 0			\qquad \mbox{(propeller is active)}~.
\end{eqnarray}

The magnetospheric radius, at which the ram pressure of the in-falling matter 
balances the magnetic pressure, is defined by,

\begin{equation}
	r_m = \left(\frac{B^2 R^6}{2\dot{M}_{_{fall}}\sqrt{2GM}}\right)^{2/7},
\label{magrad}
\end{equation}
where $B$ is the surface polar field. The in-falling matter is stopped
at around this radius.

When $B<2^{3/4}G^{1/4}M^{1/4}\dot{M}_{_{fall}}R^{-5/4}$, the magnetospheric
radius defined by Eq.(\ref{magrad}) is below the stellar surface and 
the effect of the magnetic field on the accreting matter flow is 
suppressed. The accretion flow is then assumed to be
advection dominated (we here adopt the description of Narayan \& Yi 1994,1995)
with an accretion torque,
\begin{equation}
	Q_a = \dot{M}R\left(\frac{2GM}{7R}\right)^{1/2}.
\label{torque2}
\end{equation}
We should note that the behaviour of the in-falling matter is rather uncertain
especially in the hypercritical regime in the presence of a strong magnetic 
field which possibly affects the motion of the matter.
The magnetic field may produce a strong outflow or jet, and/or destroy the
innermost structure of the in-falling flow and change it drastically. 
Modeling these flows is beyond the scope of this paper and hereafter we will
assume that the flow is described by that of a simple, advection-dominated 
disk.

When $r_m$ exceeds $R$, $Q_a$ takes the form (Watts \& Andersson 2002),
\begin{equation}
	Q_a = 2r_m^2[\Omega_k(r_m)-\Omega]\dot{M}_{_{fall}},
\end{equation}
where $\Omega$ is the rotational angular frequency of the
star, and $\Omega_k(r)$ is the Keplerian orbital frequency 
around the star at a radial distance $r$.

When the radius of the light cylinder $r_L=c/\Omega$ of the star is 
smaller than $r_m$, we also include the torque given by the magnetic dipole 
radiation;\footnote{In this case, we simply replace $r_m$ with $r_L$.}
\begin{equation}
	Q_b = -\frac{2B^2R^6\Omega^3}{3c^3}.
\label{magdipole}
\end{equation}

We adopt as the condition for the appearance of the propeller effect,
$r_m > R$ and $\Omega>\Omega_k(r_m)$.

The evolution of the system is described by the time-variation of 
$M$ and $\Omega$ (note that $R$ is constant for this polytrope).
The evolution equations are then,
\begin{equation}
	\frac{dM}{dt} = \dot{M}
\label{massevolution}
\end{equation}
\begin{equation}
	\frac{d\Omega}{dt} = \frac{Q_a+Q_b\theta(r_m-r_L)}{\kappa MR^2} 
- \frac{3}{2}\frac{\dot{M}}{M}\Omega.
\label{omegaevolution}
\end{equation}
Here $\theta(x)$ is the Helmholtz step function 
and $\kappa$ is a dimensionless constant in terms of which the moment of 
inertia is given as $I=\kappa MR^2$. For an $N=1$ polytrope $\kappa=0.261$.

Following Watts \& Andersson (2002), we start 
integrating Eqs.(\ref{massevolution}) and (\ref{omegaevolution})
from $t=t_i$ after the bounce of the collapsing core,
when a recognizable central compact object is thought to appear.
We set the total amount of in-falling debris as $\Delta M_{_{fall}}$ (typically $\sim 0.1M_\odot$).
The integration is terminated if 
$\dot{M}_{_{fall}}<10^{-4}M_\odot \mbox{yr}^{-1}$,
when the hypercritical advection dominated accretion ceases (Chevalier 1989,
Brown et al. 2000).

\section{Results}
\subsection{Evolution of rotational frequency and mass}
The parameters defining the model are the following:
1) the initial time $t_i$ at which the time integration is started 
which we choose to be $100$s (cf. Watts \& Andersson 2002);
2) the total mass of in-falling matter, $\Delta M_{fall}$ ;
3) the strength of the surface polar magnetic field $B$.
4) the initial rotational frequency of the star $f_0=\Omega(0)/2\pi$.
In what follows, we show the evolution of the mass and of the rotational 
frequency of systems with typical parameter sets.

\begin{figure}
                \centering\leavevmode
                \psfig{file=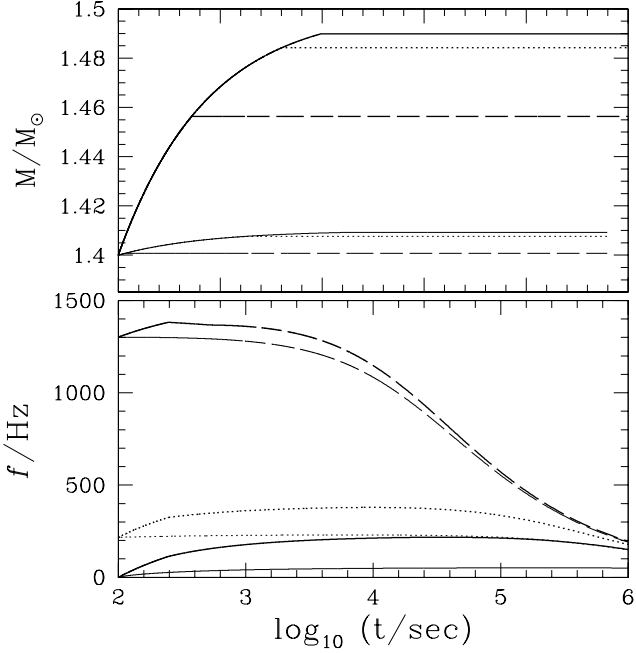,width=9cm,angle=0,clip=}
\caption{Evolution of the rotational frequency (lower panel)
and the mass (upper panel) of the model with $B=10^{14}$G 
and initial mass $M=1.4M_\odot$. 
The solid lines are for $f_0 = 0$Hz. Thin and thick lines correspond
to the values of $\Delta M_{fall}$ equal to $0.01M_\odot$ and $0.1M_\odot$,
respectively.
The dotted lines are for $f_0 = 217$Hz with the difference between the thick
and thin lines being the same as above. The dashed lines are for $f_0=1301$Hz.
}
\label{bfix14}
\end{figure}

\begin{figure}
                \centering\leavevmode
                \psfig{file=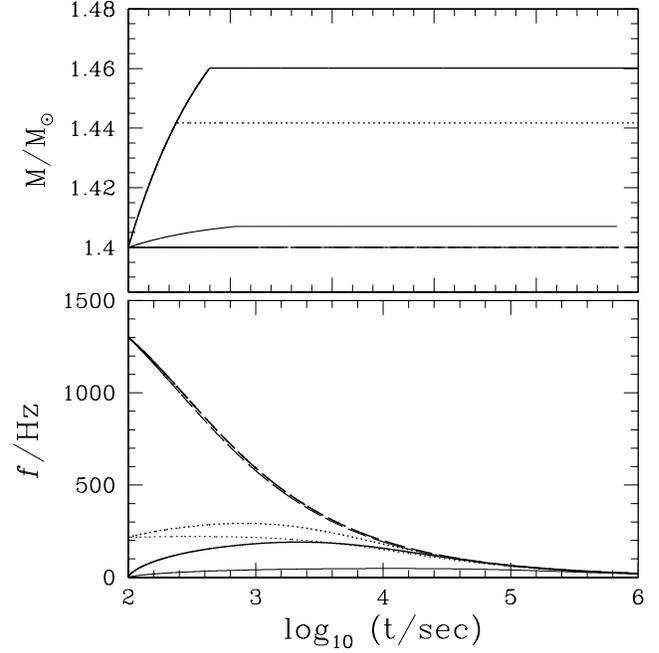,width=9cm,angle=0,clip=}
\caption{
Same as Fig.1, but with $B=10^{15}$G. The correspondence 
between line-types and parameter values are the same as in Fig.~1.
In the upper panel, the thick dashed, thin dotted and the thin dashed
lines degenerated to a single line at the bottom.
}
\label{bfix15}
\end{figure}

In Figs.~\ref{bfix14} and \ref{bfix15}, the evolutionary curves of
the mass (upper panel) and the rotational frequency (lower 
panels) are plotted.
As the in-fall rate decreases, the propeller effect becomes
active, and the matter cannot accrete onto the star. This 
occurs earlier for stronger surface magnetic fields (compare
the upper panels of Fig.\ref{bfix14} and \ref{bfix15}).
Also, the torque produced by the magnetic dipole radiation is stronger for
larger surface magnetic fields
As a result, the rotational frequency decreases faster with stronger
fields, for the same $\Delta M_{_{fall}}$.
Also note that the asymptotic behaviour of the rotational evolution
is rather insensitive to $\Delta M_{_{fall}}$.
In the later phases of the evolution, the in-falling rate is so low that
the angular frequency evolution is mainly driven by the magnetic dipole
radiation term (Eq.(\ref{omegaevolution})). Thus the dependence on the
in-fall rate can be neglected. 

Furthermore, in the later phases of evolution the frequency converges 
to the same value, irrespective of the initial rotational frequency
$f_0$.
This is also because in this phase the mass accretion rate is so small that
the evolution of the frequency is solely determined by the magnetic
dipole radiation. From Eqs.(\ref{magdipole}) and (\ref{omegaevolution}),
it is easy to see that the frequency behaves as:
\begin{equation}
	\Omega(t)^2 = \frac{1}{2\alpha (t-t_0) + \Omega(t_0)^{-2}},
\label{omegat}
\end{equation}
where $\alpha=2B^2R^4/3c^3\kappa M$ and $t_0$ is the (arbitrary) 
origin of the time.
Eq.(\ref{omegat}) shows that as $t\to\infty$, $\Omega$ becomes independent of 
$\Omega(t_0)$.


\subsection{Gravitational wave strain}
Gravitational waves from a star deformed by a magnetic field
and having its rotation axis misaligned with respect to its magnetic axis,
have been discussed by Bonazzola \& Gourgoulhon (1996).
We here use their formula to estimate the dimensionless strain of
gravitational waves.

The strain of the produced gravitational waves observed on the earth depends 
not only on the misalignment
angle of the rotational and magnetic axes, but also on the inclination
angle of the system with respect to the line of sight to the observer
(see Eqs.(20) and (21) of Bonazzola \& Gourgoulhon 1996).

The factor which does not depend on these angles is found to be (cf.
Eq.(25) of Bonazzola \& Gourgoulhon 1996):
\begin{eqnarray}
	h_{_0} &=& \frac{4G}{c^4}\frac{I\epsilon f^2}{D}\nonumber\\
		&=& 1.67\times 10^{-25} \left[\frac{f}{1\mbox{kHz}}\right]^2 \left[\frac{D}{1\mbox{Mpc}}\right]^{-1} \left[\frac{I}{10^{45}\mbox{g}\mbox{cm}^2}\right] \times\nonumber\\
&& \qquad\left[\frac{\epsilon}{10^{-6}}\right],
\label{eqh0}
\end{eqnarray}
where $f$ is the rotational frequency, $D$ is the distance to the
source, $I$ is the moment of inertia and $\epsilon$ is the
ellipticity measuring the deformation of the star. 
\footnote{Here we refer only to the fundamental frequency of the 
gravitational wave signal, i.e., the
stellar rotational frequency. There also exists an overtone whose frequency
is double of that of stellar rotation (Bonazzola \& Gourgoulhon 1996)}

Thus far we have almost no information about the ellipticity 
produced by the magnetic field, except for the upper
bounds on normal pulsars (Bonazzola \& Gourgoulhon 1996).
On the theoretical side, this is mostly due to the fact that we know 
hardly anything about the configuration of the magnetic field inside the star,
although there have been works estimating the ellipticity 
(Bonazzola \& Gourgoulhon 1996; 
Konno et al. 2000). Bonazzola \& Gourgoulhon (1996) have parametrized the
ellipticity as,
\begin{equation}
	\epsilon = \beta\frac{\mu_0}{4\pi}\frac{{\cal M}^2 R^2}{G I^2},
\end{equation}
where $\mu_0$ is the permeability of the vacuum and ${\cal M}$ is the 
magnetic dipole moment of the star. 
The parameter $\beta$ depends on the configuration of the magnetic field
in the star as well as on the compactness of the star. For a Newtonian uniform
density star with a uniform distribution of magnetic field,
$\beta = 1/5$. Bonazzola \& Goulgourhon (1996) obtained values of
$\beta$ up to $10^3$. 
On the other hand, Konno et al. (2000) studied the effect of the compactness
of the star assuming a dipolar field inside it, and found that
the ellipticity can be enhanced by general relativistic effects.

Hereafter we assume $\beta = 1$ as a 'moderate' case for the ellipticity.
As can be seen in Eq.(\ref{eqh0}), the results here can easily be scaled to
other choices for this parameter.

For simplicity, we have neglected the deformation introduced by rotation. 
If the rotation and magnetic field are weak sufficiently to be treated as a
small perturbations to the force balance of the stellar matter (a 
back-of-envelope calculation shows that the smallest magnetic field 
affecting the stellar configuration more seriously is, $B\sim 10^{17}$G),
the perturbation to $I$ can be regarded as a linear 
superposition of both these two contributions. As long as the stellar
rotation rate is not near to the mass shedding limit, the deformation by
rotation only slightly changes the moment of inertia $I$ and
has a negligible effect on the deformation which is not aligned with the 
rotation axis and which produces the gravitational radiation.

The timescale of the evolution of the rotational frequency 
is much longer than the rotational periods of the stars, so that the evolution
can be regarded as secular. In this case, the characteristic amplitude
of the signal is defined by (Owen et al. 1998),
\begin{equation}
	h_{_c} = h_{_0}\sqrt{f^2\left|\frac{dt}{df}\right|},
\label{hc}
\end{equation}
where the frequency $f$ is a secularly changing function of time.
In Fig.\ref{hc} we plot $h_{_c}$ for several parameters of our model.
The distance to the source is assumed to be $20$Mpc.
The curves labeled as A\#,B\#,C\# are for the initial rotational
frequency $f_0 = 0, 217, 1301$Hz, respectively. The number after
the letters distinguishes the strength of the magnetic field, i.e.,
``15'' corresponds to $B=10^{15}$G and ``14'' to $B=10^{14}$G. The total 
in-falling mass $\Delta M_{fall}$ is fixed to be $0.1M_\odot$.
For comparison, the noise curves of several laser interferometric
detectors (LIGO, VIRGO, GEO600) are also plotted.

For all of the curves apart from (C15), the curves consist of two parts
with increasing and decreasing amplitude $h_c$ (the lower and the upper
branches of the curves respectively). The reason for there sometimes
being two values of $h_c$ corresponding to a single value of $f$ is
that we are here calculating $h_c$ as a function of time (using Eq.(11))
and not integrating over all observed times.

In general, the evolution of the system starts with a phase of increasing
stellar rotation frequency, which is followed by a plateau of frequency
and then the frequency decreases (see eg. the lower panel of Fig.~1)
To begin with, both $h_c$ and $f$ increase; $h_c$ then rises to the maximum
value and then decreases again (with an associated decrease in $f$)
tracing out the upper branch of the curves from right to left.

%

On the other hand, for (C15), only one branch exists. This is because
in this case, the magnetic field and the initial spin frequency
are so high that the propeller mechanism is active all through the
evolution. As seen in Fig.2, the frequency decreases monotonically.

For fixed magnetic field, the upper portion of the curves converges 
asymptotically to the same line, being independent of the initial 
rotational frequency $f_0$. 
This is due to the fact that the late phase of evolution of the system 
is driven by magnetic dipole radiation, which is almost independent 
of the initial rotational frequency of the star (cf. the discussion after 
the Eq.(\ref{omegat})).


\begin{figure}
                \centering\leavevmode
                \psfig{file=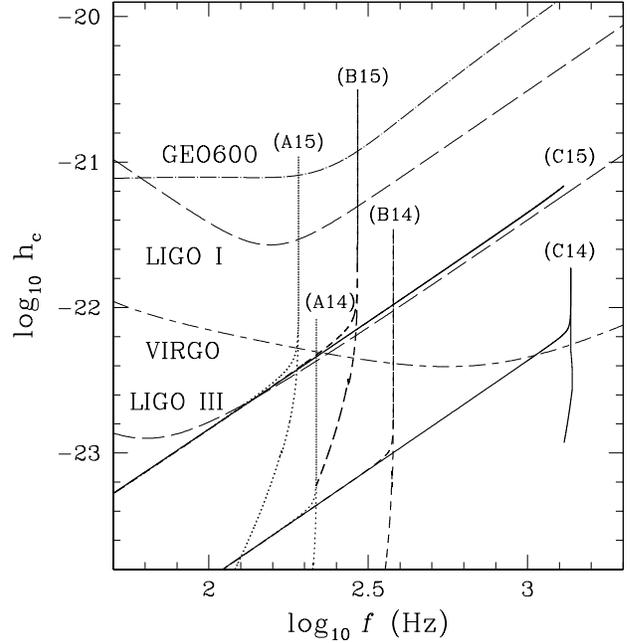,width=9cm,angle=0,clip=}
\caption{Characteristic amplitude $h_{_c}$ of sources at $20$Mpc. 
Dotted curves (A15,A14) correspond to initially non-rotating stars. (A15) is
for $B=10^{15}$G, while (A14) is for $B=10^{14}$G.
Short dashed curves (B15,B14) are those for stars 
with $f_0=217$Hz.
(B15) and (B14) are those for $B=10^{15}$G and $B=10^{14}$G, respectively.
Solid lines corresponds to those stars with the initial rotational frequency
$1301$Hz. (C15) and (C14) are those for $B=10^{15}$G and $B=10^{14}$G.
For comparison, the sensitivity curves of LIGO
detectors (long-dashed lines; reproduced by the fitting 
formula by Owen et al. (1998)), 
that of VIRGO detector (short- and long-dashed line; fitting formula 
is obtained by J.Y.~Vinet ({\tt http://www.virgo.infn.it/})),
and that of GEO600 (dot and short-dashed line; fitting formula
are found in Damour et al. 2001) are plotted.}
\label{hc}
\end{figure}	

\begin{figure}
                \centering\leavevmode
                \psfig{file=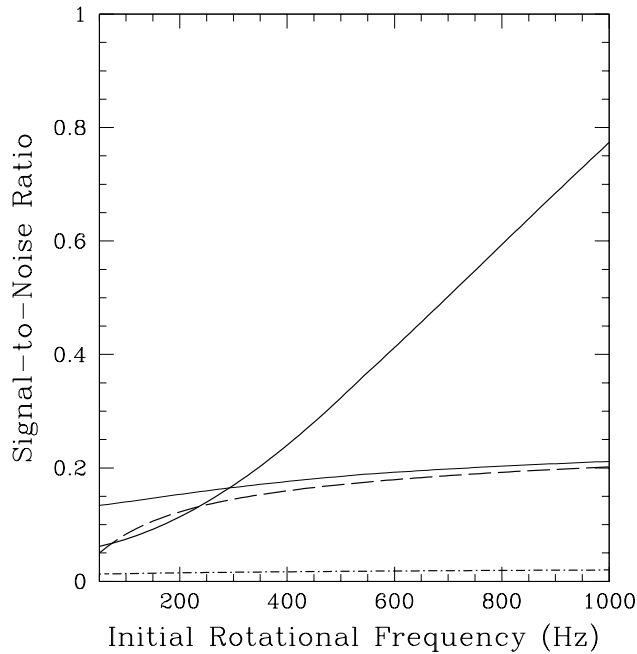,width=9cm,angle=0,clip=}
\caption{Signal-to-Noise ratio for sources at $20$Mpc plotted 
against the initial rotational frequency of the stars. 
The thick solid line corresponds to that of VIRGO in the case of
$B=10^{15}$G and $\Delta M_{fall}=0.1M_\odot$.
The thin solid line is that of LIGO III for the same star.
The dashed line is that of VIRGO for a star with 
$B=10^{15}$G and $\Delta M_{fall}=0.01M_\odot$, 
whereas the dotted-dashed
line corresponds to the result of VIRGO for a star with
$B=10^{14}$G and $\Delta M_{fall}=0.1M_\odot$.}
\label{sn}
\end{figure}

To see the detectability of the sources, the signal-to-noise ratio ($S/N$)
is computed using the formula (Owen et al. 1998);
\begin{equation}
	\frac{S}{N} = \sqrt{2\int_{f_{_{\rm min}}}^{f_{_{\rm max}}} 
	\frac{df}{f} 
	\left(\frac{h_{\rm c}}{h_{\rm rms}}\right)^2}\quad,
\end{equation}
where $f_{_{\rm min}}$ and $f_{_{\rm max}}$ are the minimum and the maximum
frequencies of the source during the observation and $h_{\rm rms}$ is the noise
spectrum of the detector. We take the approximate formulae for
the noise of VIRGO (by J.Y.~Vinet; {\tt http://www.virgo.infn.it/})
and LIGO III (Owen et al. 1998).
In Fig.\ref{sn}, 
the value of computed $S/N$ is plotted as a function of the initial rotation
frequency for a star at a distance of $20$Mpc.
As expected, the $S/N$ becomes larger with increases of 
the stellar magnetic field,
the total mass of the in-falling matter and the initial rotation 
frequency of the star. 
With the parameters adopted here, we conclude that sources at this
distance cannot be observed by LIGO III or VIRGO.
However, as mentioned before, the parameter $\beta$ which measures
the deformation of the star by the magnetic field is quite uncertain.
We should note again that we have here adopted a moderate choice for this 
parameter, $\beta = 1$. For instance, if this value goes up by one 
order of magnitude, an initially rapidly rotating ($f\sim 500$Hz), 
highly magnetized ($B\sim 10^{15}$G) star can be observed at $S/N\sim 3$.

\section{Summary and comments on assumptions}
Using a simple model, we have computed the evolution of highly 
magnetized neutron stars with hypercritically in-falling matter
just after their birth. The presence of a strong magnetic field
would introduce a deformation of the star. If the axis of deformation 
were misaligned with the rotation axis, the object would emit
gravitational waves with frequencies equal to the rotational one
and its overtone.
We have computed the characteristic amplitude of gravitational waves 
produced in this way.
The results obtained indicate that if a strong magnetic field 
($B\sim 10^{15}$G)
efficiently deforms the star ($\beta\sim 10$),
we could observe the resulting signal from a source in the Virgo Cluster,
when matched-filtering techniques are used in up-coming detectors.


Our model is built on several assumptions. Two of them in particular
should be commented on.

Firstly, 
we assume that the accretion flow can be well-described by
a simple advection dominated disk. In the present context, the accretion
rate is such that an accretion shock is developed far outside the stellar
surface (Armitage \& Livio 2000; see eq.(17) of Brown et al. 2000).
Internal to the shock, the accretion flow can be again hypercritical.
However, it is possible that the system might have a jet or
an explosive outflow to remove the matter in-flowing onto the
central objects (Armitage \& Livio 2000). In that case, the accretion
rate onto the object would be significantly reduced. 
At present the mechanism of formation of
an outflow from an accretion disk is unknown.
For instance, the standard magnetically driven outflow (Blandford 1976; 
Lovelace 1976) may not work because the ram pressure of the inflow here is 
comparable with or exceeding that of the magnetic field of the central star.

Secondly we assume that the magnetic field inside and at the 
surface of the star does not change in the course of the evolution 
of the system.
The decay of neutron star magnetic fields has been an important but
unsolved issue which is related to the observational data for radio 
pulsars and low mass X-ray binaries (Possenti 1999).
To explain the different strengths of the magnetic field in 'younger'
systems (ordinary radio pulsars) and 'older' ones (low mass X-ray binaries
or millisecond pulsars), some kind of mechanism for field decay seem
to be needed. The 'accretion driven decay' model and 
the 'spin driven decay' model (Possenti 1999) might both be relevant for
our model system here although their applicability is not clear since they
have much stronger field. Also notice that the time 
scale of the decay in these scenarios should be much longer than the time 
scale of the evolution of the system (see also Zanotti \& Rezzolla 2001). 
\section*{Acknowledgement}
The author thanks M. Colpi, K. Konno, J. Miller, L. Rezzolla, B.S. Sathyprakash and A. Watts for useful suggestions and discussions.
This research has been supported by MIUR and the EU Programme 'Improving the Human Research Potential and the Socio-Economic Knowledge Base' (Research
Training Network Contract HPRN-CT-2000-00137).

\end{document}